\documentclass[twocolumn]{aastex6}
\usepackage{color}
\usepackage{appendix}

\newcommand{\ha}{\hbox{H$\alpha$}}
\newcommand{\hb}{\hbox{H$\beta$}}

\newcommand{\nii}{\hbox{N\,{\sc ii}}}
\newcommand{\gsim}{\lower.5ex\hbox{$\; \buildrel > \over \sim \;$}}
\newcommand{\lsim}{\lower.5ex\hbox{$\; \buildrel < \over \sim \;$}}
\newcommand{\oi}{\hbox{[O\,{\sc i}]}}
\newcommand{\oii}{\hbox{[O\,{\sc ii}]}}
\newcommand{\oiii}{\hbox{[O\,{\sc iii}]}}

\newcommand{\sii}{\hbox{[S\,{\sc ii}]}}

\newcommand{\feii}{\hbox{[Fe\,{\sc ii}]}}
\collaborationName{Friends of AASTeX}

\begin{document}


\title{B2 0003+38A: a classical flat-spectrum radio quasar hosted by a rotation-dominated galaxy with a peculiar massive outflow}


\author{Qinyuan Zhao  \altaffilmark{1,2}, Luming Sun \altaffilmark{3} , Lu Shen \altaffilmark{1,2}, Guilin Liu \altaffilmark{1,2} , Hongyan Zhou \altaffilmark{1,2,4} , Tuo Ji \altaffilmark{2,4}}
\affil{$^{1}$ Key Laboratory for Research in Galaxies and Cosmology, Department of Astronomy, University of Science and Technology of China, Chinese Academy of Sciences, Hefei, Anhui 230026, China; zqy94070@mail.ustc.edu.cn, lmsun@ustc.edu.cn, glliu@ustc.edu.cn, zhouhongyan@pric.org.cn\\
$^{2}$ School of Astronomy and Space Sciences, University of Science and Technology of China, Hefei 230026, China\\
$^{3}$ Department of Physics, Anhui Normal University, Wuhu, Anhui, 241002, China\\
$^{4}$ Antarctic Astronomy Research Division, Key Laboratory for Polar Science of the State Oceanic Administration, Polar Research Institute of China, Shanghai, China}
\thanks{corresponding authors: Luming Sun, Guilin Liu, Hongyan Zhou}
\begin{abstract}
 
We present a detailed analysis of the single-slit optical spectrum of the Flat-Spectrum Radio Quasar (FSRQ)  B2 0003+38A, taken by the Echellette Spectrograph and Imager (ESI) on the Keck II telescope. This classical low-redshift FSRQ  ($z=0.22911$, as measured from the stellar absorption lines)  remains underexplored in its emission lines, though its broad-band continuum properties from radio to X-ray is well-studied.  After removing the unresolved quasar nucleus and the starlight from the host galaxy, we obtain a spatially-resolved 2-D spectrum, which clearly shows three components, indicating a rotating disk, an extended emission line region (EELR) and an outflow. The bulk of the EELR, with a characteristic mass $M_{\rm EELR}\sim 10^{7}~\rm M_{\odot}$, and redshifted by $v_{\rm EELR}\approx 120$ km s$^{-1}$ with respect to the quasar systemic velocity, shows a one-sided structure stretching to a projected distance of $r_{\rm EELR}\sim 20$ kpc from the nucleus. The rotation curve of the rotating disk is well consistent with that of a typical galactic disk, suggesting that the FSRQ is hosted by a disk galaxy. This conclusion is in accordance with the facts that strong absorption in the HI 21-cm line was previously observed, and that Na I$\lambda\lambda5891,5897$ and Ca II$\lambda\lambda3934,3969$ doublets are detected in the optical ESI spectrum. B2 0003+38A will become the first FSRQ discovered to be hosted by a gas-rich disk galaxy, if this is confirmed by follow-up deep imaging and/or IFU mapping with high spatial resolution. These observations will also help unravel the origin of the EELR.

\end{abstract}



\section{Introduction} 
\label{sec:introduction}


Blazars are a class of rare radio-loud Active Galactic Nuclei (AGN) \citep{Sandage1965}, characterized by their flat radio spectra, rapid variability in multiwavelength emission, significant polarization, and bi-modal synchrotron/Compton spectral energy distributions (SEDs). These characteristics are likely consequences of shocks driven by a powerful relativistic jet pointing to a direction close to our line of sight (l.o.s) and roughly perpendicular to the accretion disc \citep{Urry1995}. Blazars may show strong broad emission lines in their optical spectra, similar to typical quasars, or be featureless in their optical spectra; these blazars are called Flat-Spectrum Radio Quasars (FSRQs) and BL Lacertae objects (BL Lacs), respectively.



Observations show that the majority of blazars, regardless of their type, are hosted by giant elliptical galaxies (for BL Lacs, e.g.,\citealt{Stickel1991,Kotilainen1998,Kotilainen1998a,Falomo1999,Urry2000,Falomo2000,Kotilainen2005,Hyvoenen2007,LeonTavares2011,Falomo2014}; for FSRQs, e.g.,\citealt{OlguinIglesias2016}). \citet{OlguinIglesias2016} presented deep NIR images of a sample of 19 ($0.3 < z < 1.0$) FSRQs, finding that the host galaxies of their sample are luminous and apparently to follow the $\mu_{e}-R_{eff}$ relation for ellipticals and bulges, consistent with the conclusion based on BL Lacs \citep{Stickel1991}.

As yet, whether a blazar can be hosted by a disk galaxy remains an open question. Only a handful of cases have been reported that BL Lacs are found to be hosted by disk-dominated galaxies (e.g.,\citealt{Halpern1986,Abraham1991,Wurtz1996}). \citet{Nilsson2003} analyzed 100 BL Lacs from the ROSAT-Green Bank (RGB) sample obtained by using the Nordic Optical Telescope (NOT), finding that all of their spatially resolved objects are better fitted by an elliptical galaxy model ($\beta$ = 0.25) than by a disk galaxy model ($\beta$ = 1.0), though with two exceptions that may be hosted by disk galaxies (\citealt{Abraham1991,Wurtz1996} for 1419+543 and \citealt{Urry1999} for 1540+147), whose bulk properties, however, are indistinguishable from normal elliptical galaxies.
The host of the two BL Lacs, 1415+255 and 1413+135, were originally identified to be disk-dominated galaxies (\citealt{Stocke1992,Halpern1986,Lamer1999}, respectively), but recent works, based on imaging with higher resolution, classify 1415+255 as an isolated giant elliptical galaxie \citep{Gladders1997}, highlighting the importance of high quality data. Furthermore, \citet{Scarpa2000} investigated 69 spatially-resolved BL Lacs hosts and found one case of disk (that of 0446+449) and two cases with disk models preferred (1418+546, 0607+711). Despite the efforts that have been made, a definitive evidence showing that a blazar can be hosted by a disk galaxy (especially for FSRQs) remains to be found. This is mostly due to the lack of high-resolution imaging or specially resolved spectra, and the brightness of blazars is so high in optical and IR bands that the hosts are outshined.

A combination of high-resolution imaging and spatially-resolved spectroscopy may shed light on the morphology and dynamics of the host galaxy. Hence, we conduct a case study on a confirmed FSRQ at $z$ = 0.229, known as B2 0003+38A (a.k.a. S4 0003+38 or J0005+3820) using the long-slit optical spectroscopy taken by the Echellette Spectrograph and Imager (ESI) at the Keck II observatory. This paper is structured as follows: after an overview of B2 0003+38A in previous studies, we describe observations and data reduction in Section 2. In Section 3, we present the analysis of the spectral data. In Section 4, we discuss the gas kinematics in the host galaxy and the extended emission line region. We conclude with a summary in Section 5. Throughout this paper, we adopt a cosmology with $H_0$ = 0.7 km s$^{-1}$ Mpc$^{-1}$, $\Omega_{m} = 0.3$, $\Omega_{\Lambda} = 0.7$. The wavelengths of all the spectral lines are given in vacuum.

\section{Observation and Data Reduction} 
\label{sec:section_name}


\subsection{The FSRQ B2 0003+38A}  \label{sec:dr}

B2 0003+38A is classified as an FSRQ due to its broad emission lines in the optical spectrum and the flat radio spectrum with $\alpha = -0.3$ \citep{Stickel1994,Healey2007,Massaro2009}. Resolved by a VLBI mapping at 2.29 GHz \citep{Morabito1982}, it shows a prominent core that emits more than 95 per cent of the 2.3 GHz flux, along with a weak jet pointing to the sourth-east \citep{Fey2000}, as shown in Figure ~\ref{fig:fig0}. Kinematic analyses show quasi-stationary knots at the jet base and relativistic motions downstream \citep{Hervet2016}. \citet{Aditya2018} detected HI 21-cm absorption towards this source, finding the velocity-integrated HI 21-cm optical depths to be 1.943 $\pm$ 0.057 km s$^{-1}$, and the HI column density to be 3.54 $\pm$ 0.11 $\times 10 ^{20}$ cm$^{-2}$.

\subsection{Optical Long-slit Spectrometry}  \label{sec:dr}

\begin{figure}[htb]
\centering
\includegraphics[width=0.45\textwidth]{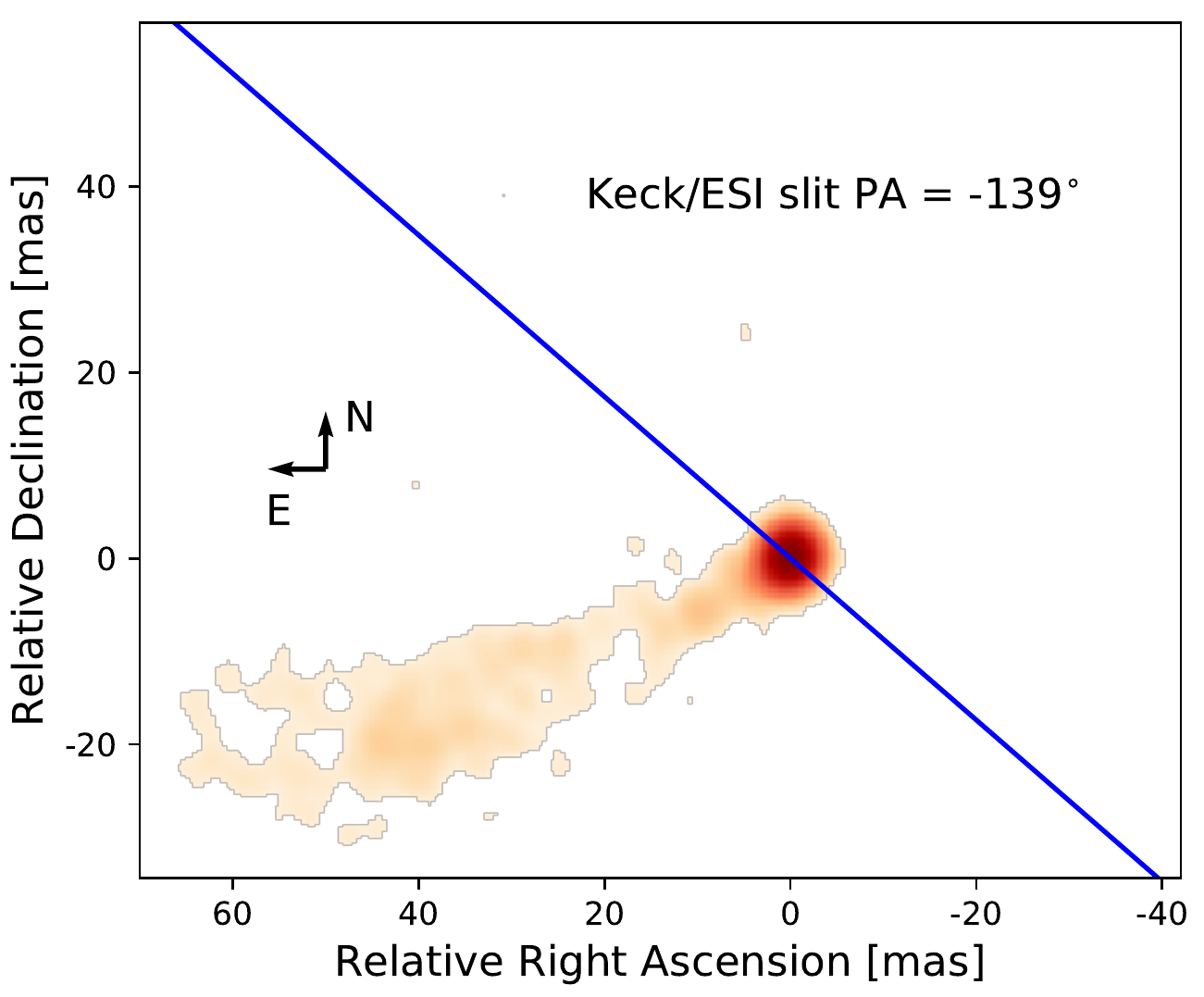}
\caption{The VLBI 2.292GHz cutout centered on the B2 0003+38A  \citep{Lister2013}. The orientation of the ESI/Keck slit is illustrated by the blue line.  The 1.25$\arcsec$ slit width is larger than the size of the cutout. The slit covers the quasar and jet. }

\label{fig:fig0}
\end{figure}

The quasar B2 0003+38A was observed by Keck-II Echellette Spectrograph and Imager (ESI) long-slit spectrography in its echelle mode on 2015 Sep 9.
The spectrum was taken in the optical band (wavelength coverage $\lambda$ $\sim$ 3995 $\mathrm{\AA}$ - 10198 $\mathrm{\AA}$ in the observer's frame), covering 3250 $\mathrm{\AA}$ to 8297 $\mathrm{\AA}$ in the rest frame for our target at $z \sim 0.2$.
A 1.25$\arcsec$-wide slit was employed, resulting in an instrumental dispersion $\sigma$ of 22 km s$^{-1}$.
The position angle of the slit is 41 deg (north to east, Figure ~\ref{fig:fig0}). The slit width allows for both the blazar and its jet to be covered.
Two exposures were taken with an integration time of 10 minutes per each.
Arc lamps were used during the observing campaign for wavelength calibration, and the standard star BD+28 4211 was observed about one hour before the target with identical settings for the purpose of flux calibration.

\subsection{Data reduction}  \label{sec:dar}


We use a customized routine based on the IDL package \textbf{XIDL}\footnote{http://www2.keck.hawaii.edu/inst/esi/ESIRedux/index.html} to obtain a two-dimensional (hereafter 2-D) spectrum.
Our data reduction consists of four steps:
(1) after bias corrections, flat-fielding and removing cosmic rays, we stack the two exposures;
(2) the nucleus of the target quasar is traced along 10 echelle orders; and for each order, we resample the 2-D spectrum so that a resultant pixel corresponds to 10 km s$^{-1}$ in the wavelength direction and 0.15$\arcsec$ in the spatial direction;
(3) sky subtraction and flux calibration are applied on these resampled 2-D spectra;
(4) we combine these resampled 2-D spectra from the 10 orders.

\section{Data analysis and results} 
\label{sec:section_name}

We conduct kinematic analyses using both the 1-D and 2-D spectrum to investigate the gas motion in the host galaxy. In Section \ref{sec:nsa}, we perform a model fitting to the 1-D spectra to contextualize the quasar and stellar components, and the information that the fitting delivers is discussed in Section \ref{sec:2ds}. With the quasar and stellar contributions removed, we scrutinize the spatially resolved emission lines in the 2-D spectra,  and the gas kinematics is described in Section \ref{sec:2dsnl}. We observe three components exist in the gas: a rotating component, an extended emission-line component, and an outflow component. The former two components are further analyzed in Sections \ref{sec:disk} and \ref{sec:eelr}.


\subsection{Spectrum of the nucleus}  \label{sec:nsa}

\begin{figure*}[htb]
\centering
\includegraphics[width=0.95\textwidth]{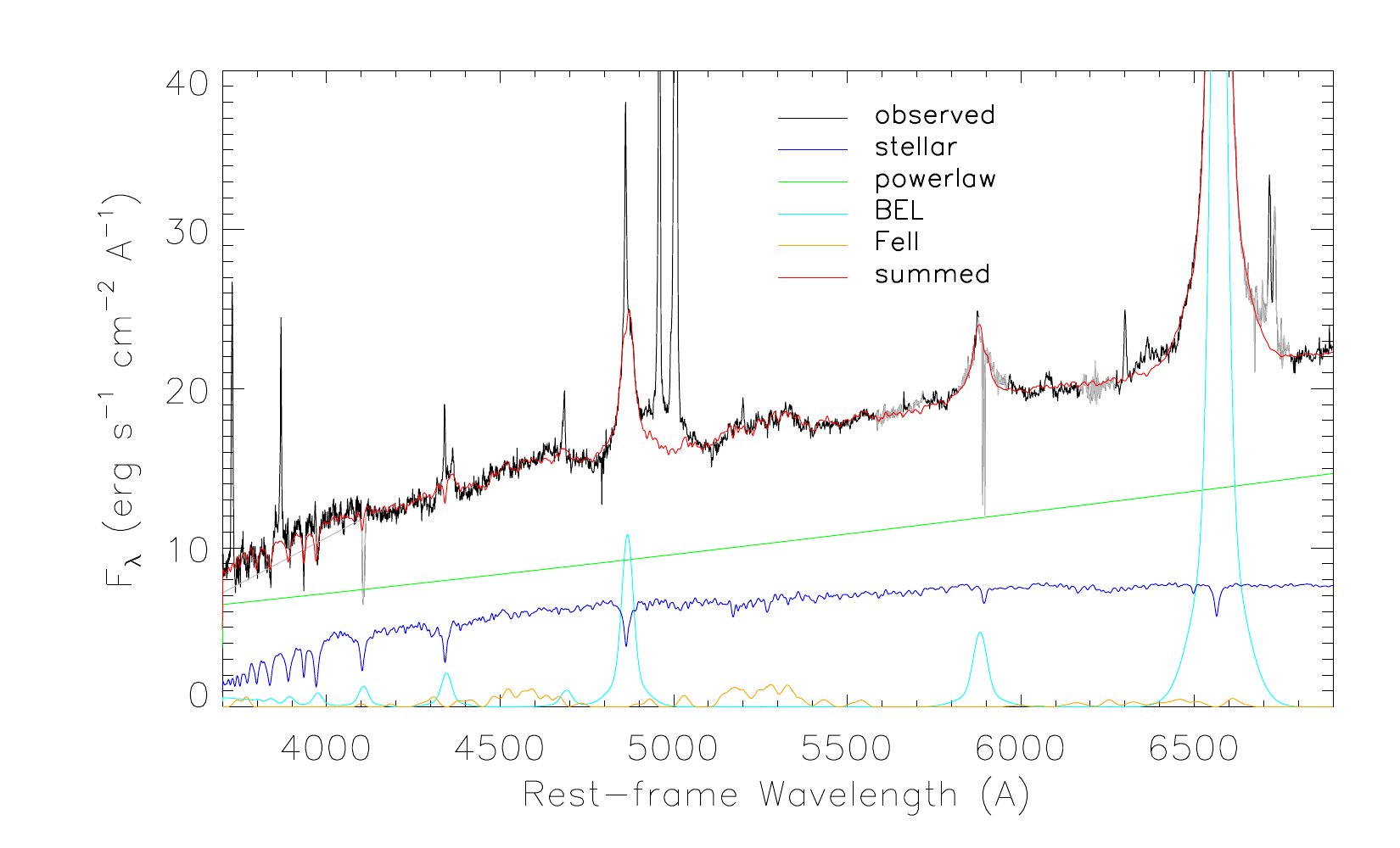}
\caption{The observed nuclear spectrum (black) and the best-fitted spectrum(red) are shown in regions where the NELs affect little. The four components of the best-fitted model are shown in different colored lines: the stellar component in blue, the power-law component in green, the BEL component in cyan and the Fe II component in yellow. The spectrum in regions affected by the telluric absorption is plotted in grey line.}

\label{fig:fig1}
\end{figure*}

We extracted the one-dimensional (hereafter 1-D) spectrum from the 2-D spectrum as a result of data reduction by applying an aperture with a size of 3$\arcsec$ $\times$ 1.25$\arcsec$ centered on the quasar nucleus, as shown in Figure ~\ref{fig:fig1} (black lines). AGNs features, including broad emission lines (BELs), narrow emission lines (NELs) and Fe II bumps, are seen therein. However, the existence of high-order Balmer absorption lines indicate that the continuum contribution from starlight.
To disentangle different contributions to the spectrum, we construct a model consisting of four components: a stellar, a power-law, a BEL and an Fe II component.\\
1. The stellar component: we assume a single simple stellar population (SSP), utilizing the library of \citet{Bruzual2003}. When fit to the data, the template is shifted to a redshift $z_\star$, broadened by convolving to a Gaussian parameterized by a velocity dispersion parameter $\sigma_\star$, and multiplied by the extinction curve of the Small Magellan Cloud (SMC) parameterized by $E(B-V)_\star$ \citep{Pei1992}.\\
2. The power-law component: we assume a formulation of $f_\lambda(\lambda)=C \lambda^\alpha$, where $\alpha$ is the exponent and $C$ is the normalization.\\
3. The BEL component: for this component, we mainly account for the Balmer series, He I $\lambda$5876 and He II $\lambda$4686. We assume that the BEL profiles can be represented by a linear combination of two Gaussians, whose mean and standard dispersion are fixed but the flux is allowed to vary. Furthermore, we fix the flux ratios of higher-order Balmer BELs to H$\gamma$ BEL to those under the ``Case B" situation \citep{Storey1995}, assuming a typical circumstance for broad line regions with $T_e=15,000$ K and $n_e=10^9$ cm$^{-3}$.\\ 
4. The Fe II component: we employ the Fe II template constructed by \citet{VeronCetty2004}.
After experiments with various combinations of different Fe II lines, we conclude that only narrow Fe II lines are necessary for fitting the observed Fe II complex. We further assume that these different narrow Fe II lines can be represented by single Gaussians with a fixed redshift and a fixed width. The flux ratios between different Fe II lines are taken from the template given in \citet{VeronCetty2004}. In addition, we assume the SMC extinction curve for the dust attenuation and reddening of the Fe II component parameterized by $E(B-V)_{\rm FeII}$.\\
We then fit the spectrum of the nuclear region, where the contamination from NELs is minimal, resulting in a minimized reduced Chi-square of 1.99. This fit can be further improved if the regions affected by telluric absorptions are dismissed (reduced $\chi^2$ = 1.66, Figure ~\ref{fig:fig1}). The individual components of the best-fit spectra are depicted by colored lines in the same figure, and the best-fit parameters are tabulated in Table ~\ref{tab:tab1}.

\begin{table}[htb]\footnotesize
\caption{Parameters for modelling the nuclear spectrum.}
\begin{tabular}{ccc}
\hline
\hline
Parameter   & Value & Unit\\
\hline
\multicolumn{3}{c}{stellar component}\\
\hline
age                & $450$ & Myr \\
$M_V$              & -22.77 & mag \\
$z_\star$          & $0.22911\pm0.00001$ \\
$\sigma_\star$     & $150\pm20$ & km s$^{-1}$ \\
$E(B-V)_\star$     & $0.62\pm0.02$ & mag\\
\hline
\multicolumn{3}{c}{power-law component} \\
\hline
$f_{\lambda5100}$  & $9.84\pm0.06$ & $10^{-17}$ erg s$^{-1}$ cm$^{-2}$ \AA$^{-1}$ \\
$\alpha$           & $1.32\pm0.02$ & \\
\hline
\multicolumn{3}{c}{BEL component} \\
\hline
$\Delta v_1$       & $263\pm3$   & km s$^{-1}$ \\
$\sigma_1$         & $974\pm4$   & km s$^{-1}$\\
$\Delta v_2$       & $250\pm10$  & km s$^{-1}$\\
$\sigma_2$         & $2960\pm20$ & km s$^{-1}$\\
frac$_1$           & $0.598\pm0.003$ & \\
$f_{\rm H\alpha}$  & $4740\pm20$ & $10^{-17}$ erg s$^{-1}$ cm$^{-2}$\\
$f_{\rm H\beta}$   & $587\pm3$   & $10^{-17}$ erg s$^{-1}$ cm$^{-2}$\\
$f_{\rm H\gamma}$  & $103\pm4$   & $10^{-17}$ erg s$^{-1}$ cm$^{-2}$\\
$f_{\rm HeI5876}$  & $307\pm3$   & $10^{-17}$ erg s$^{-1}$ cm$^{-2}$\\
$f_{\rm HeII4686}$ & $55\pm3$    & $10^{-17}$ erg s$^{-1}$ cm$^{-2}$\\
\hline
\multicolumn{3}{c}{Fe II component}\\
\hline
$\Delta v$         & $430\pm20$  & km s$^{-1}$\\
$\sigma$           & $720\pm20$  & km s$^{-1}$\\
$f_{\lambda4590}$  & $1.7\pm0.2$ & $10^{-17}$ erg s$^{-1}$ cm$^{-2}$ \AA$^{-1}$\\
$E(B-V)_{\rm FeII}$& $0.19\pm0.04$ & mag \\
\hline
\end{tabular}
\label{tab:tab1}
\end{table}

\subsubsection{A Narrow Absorption Line System} \label{sec:nabsya}

\begin{figure}[htb]
\centering
\includegraphics[width=0.5\textwidth]{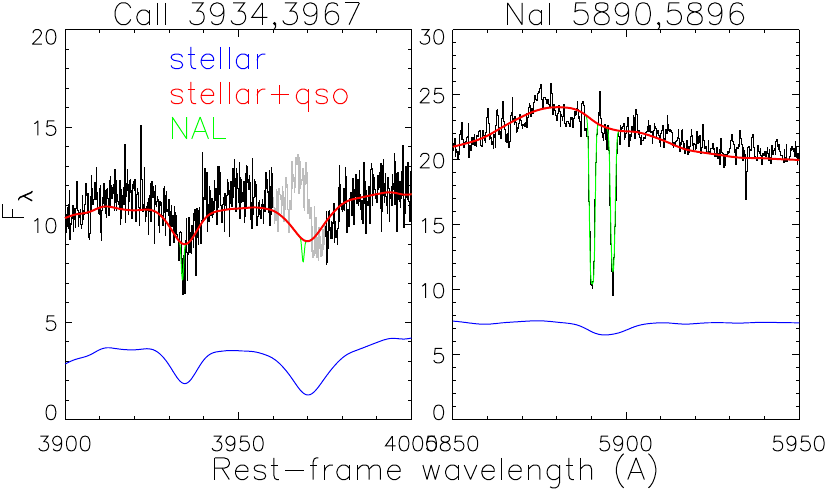}

\caption{ The observed spectrum near the Ca II (left) and Na I (right) narrow absorption lines are shown in black lines. The best-fitted spectra are shown in colored lines based on their components. The best-fitted spectra of the stellar component and stellar plus quasar components, and deep narrow absorption lines are shown in blue, red, and green, respectively.
\label{fig:fig2}}
\end{figure}

The nuclear spectrum of B2 0003+38A shows deep and narrow Na I $\lambda\lambda$5890,5896 absorption lines ( Figure ~\ref{fig:fig2}), corresponding to a redshift of $\sim$0.22883, which is about 70 km s$^{-1}$ blueshifted relative to the stellar redshift (0.22911).
Unlikely to originate from the stellar populations, these are likely absorption lines from absorbers that happen to lie in our line of sight towards the quasar nucleus.
Narrow Ca II $\lambda$3934 absorption is detected at the same redshift as that of NaI absorption, implying for the same absorption system, though we cannot affirm the existence of the Ca II $\lambda$3969 absorption line due to the influence of [NeIII] $\lambda$3869 and H$\epsilon$ emission lines. Assuming that the starlight is not affected by the absorber mentioned above, we consider the case of partial coverage:
\begin{equation}
f(\lambda) = f_{\rm qso}(\lambda) \times (1-C_f+C_f \times e^{-\tau(\lambda)}) + f_{\rm stellar},
\end{equation}
where $f_{\rm qso}$ is the flux density of the emission from the quasar nucleus (including the power-law continuum, the BEL and the FeII components), $f_{\rm stellar}$ is the flux density of stellar emission, $C_f$ is the covering factor, and $\tau(\lambda)$ is the optical depth of absorption.
The $\tau(\lambda)$ profile that we adopt for each absorption line is Gaussian with a fixed velocity dispersion. We perform this nuclear spectrum fitting in the vicinity of the narrow absorption lines, and the resultant best-fit model and the corresponding parameters are given in Figure ~\ref{fig:fig2} and Table ~\ref{tab:tab2}, respectively.
We find the redshift of these absorption lines to be $z = 0.228827\pm0.000002$, highly consistent with that of the H I 21-cm absorption ($z\sim0.2288$; \citealt{Aditya2018}). 
In addition, the H I 21-cm absorption line is spectrally unresolved, in line with the narrowness of Na I and Ca II absorption lines ($\sigma\sim23$ km s$^{-1}$).
These consistencies strongly imply for a relation between the absorption of Na I/Ca II and that of H I.
\citet{Aditya2018} measured the H I column density to be $(3.54\pm0.11)\times (T_s/100\ {\rm K})\times 10^{20}$ cm$^{-2}$, where $T_s$ is the spin temperature.
The H I column density $N({\rm H})$ and the Na I column density $N({\rm Na I})$ are related by the following equation (e.g., \citealt{Rupke2005}):
\begin{equation}
N({\rm H}) = N({\rm Na I}) (1-y)^{-1} 10^{-(a+b)},
\end{equation}
where $y$ is the ionization fraction, $a$ is the Na abundance, and $b$ describes the depletion onto dust.
If we adopt $y=0.9$ following \citet{Rupke2005}, a solar abundance (so that $a=-5.69$), and the canonical Galactic depletion value of $b=-0.95$, then we see that the measured $N({\rm Na I})$ corresponds to $N({\rm H}) \sim$ 1.1$\times10^{21}$ cm$^{-2}$ , a value close to the result of radio spectral analysis.

\begin{table}[htb]\footnotesize
\caption{Parameters for modelling the narrow absorption lines.}
\begin{tabular}{ccc}
\hline
\hline
Parameter   & Value & Unit\\
\hline
z              & $0.228827\pm0.000002$ & \\
$\sigma$       & $23.4\pm0.6$ & km s$^{-1}$ \\
$C_f$          & 0.78         & \\
$N_{\rm NaI}$  & $26\pm2$     & $10^{12}$ cm$^{-2}$ \\
$N_{\rm CaII}$ & $3.4\pm0.7$  & $10^{12}$ cm$^{-2}$ \\
\hline
\end{tabular}
\label{tab:tab2}
\end{table}

\subsection{The nature of the quasar}  \label{sec:2ds}

BELs in the optical spectrum are unambiguous emitted from the quasar nucleus.
We measure the Balmer decrement of the BELs to be $8.07\pm0.05$, significantly higher than the theoretical ``Case" B value of 2.7 \citep{Gaskell2017}, indicating heavy dust reddening towards the quasar nucleus.
Assuming an SMC extinction curve results in an estimation that $E_{\rm B-V} \sim 1.2$.
The power-law component contributes 60\% to 70\% of the total continuum flux in the wavelength range of 4000-7000 \AA.
The quasar continuum is remarkably red with a power-law index $\alpha$ of $1.32\pm0.02$, probably a result of the synchrotron emission from the radio jet, and/or the reddened thermal emission from quasar nucleus.


In view of the heavy dust extinction in optical bands, we use the infrared continuum to estimate the bolometric luminosity of the quasar. As the first step, we obtain the 5$\mu$m monochromatic luminosity using the infrared spectral energy distribution constructed from the ALLWISE photometry of the quasar, finding that $\nu L_\nu$(5$\mu$m) $= 8\times10^{44}$ erg s$^{-1}$. We then follow \citet{Richards2006} to apply a bolometric correction factor of 8, reaching a bolometric luminosity of $6\times10^{45}$ erg s$^{-1}$, though this value may have been overestimated due to the possible contribution of synchrotron emission from the jet at 5 $\mu$m.

\subsection{2-D Spectra of Narrow Emission Lines and Spatial Decomposition}  \label{sec:2dsnl}

\begin{figure*}[htb]
\centering
\includegraphics[width=0.9\textwidth]{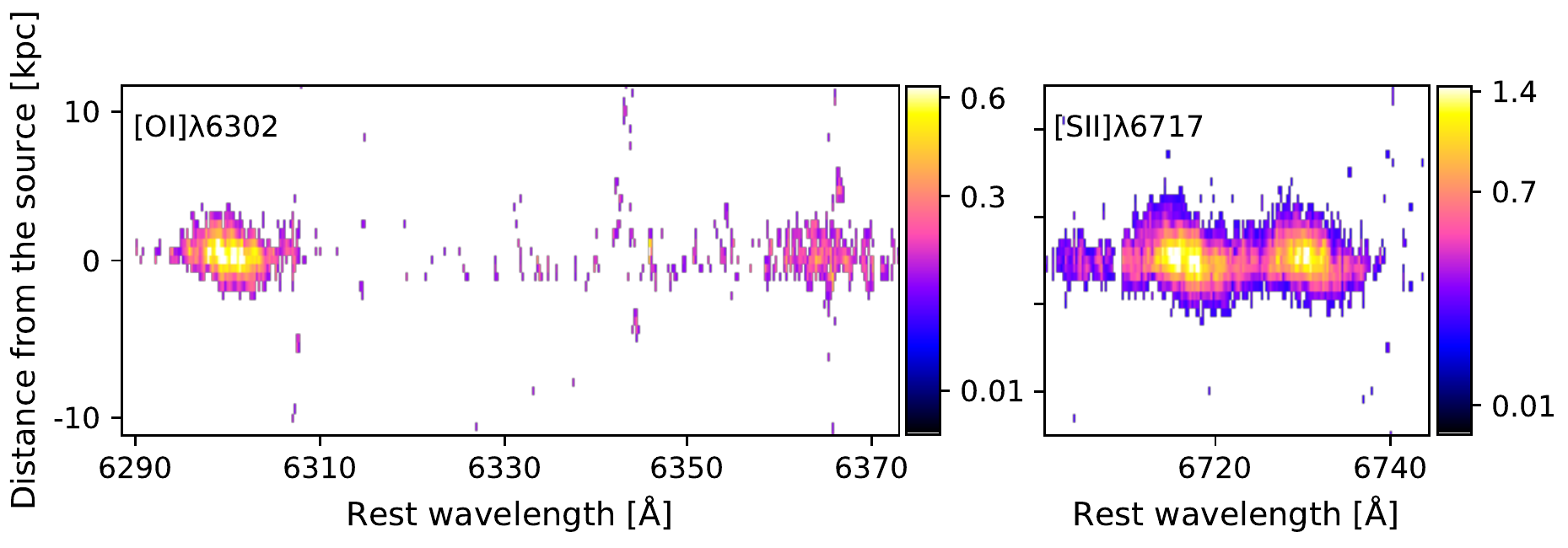}
\includegraphics[width=0.9\textwidth]{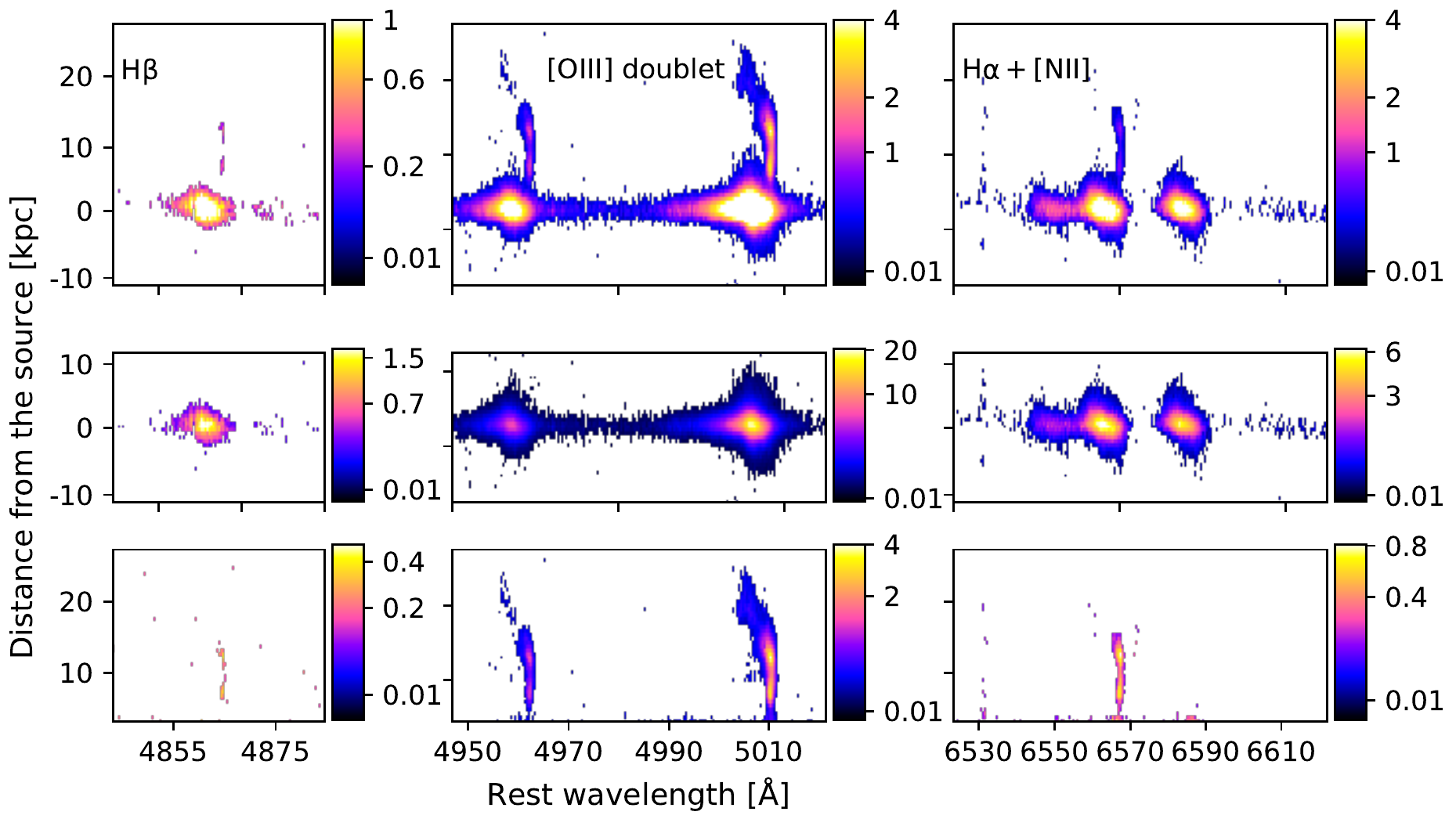}

\caption{Top two rows: the observed 2-D spectra of the \oi, \sii, \hb, \oiii\ and \ha\ after the continuum and  \feii\ subtractions. Only spaxels where the flux is detected with S/N $>$ 4 are plotted. The color of spaxels are scaled by flux density (squared scale, in units of 10$^{-17}$ erg s$^{-1}$ cm$^{-2}$ arcsec$^{-2}$ $\rm \AA^{-1}$). The spectral direction is horizontal, and the spatial direction is vertical (southwest is in positive). The 2-D spectra of \oi- and \sii-emitting NELs show a velocity gradient across the spatial extent, indicating a rotation-dominated disk. The 2-D spectra of \ha\ and \hb\ NELs  imply the existence of an extended region located $\sim$4--20 kpc southwest to the quasar nucleus, whose position and velocity is reminiscent of extended emission line regions (EELRs) around quasars. The 2-D spectra of \oiii\ NEL is more complicated: besides the two components mentioned above, an additional blue-shifted component centered on the quasar nucleus is evident. Such a profile is conventionally considered to be suggestive of the existence of outflowing gas. Our multiple-Gaussian decomposition of these emission lines allows us to isolate the contribution from these components. Third row: the reconstructed 2-D spectra of blue peaks colored by the flux density. This shows the rotating disk plus the outflow component. Bottom row: the reconstructed 2-D spectra of red peaks colored by the flux density. This reveals the EELR extended to Southwest $\sim$ 20 kpc.  
\label{fig:oisii}}
\end{figure*}

\begin{figure*}[htb]
\centering

\includegraphics[width=0.9\textwidth]{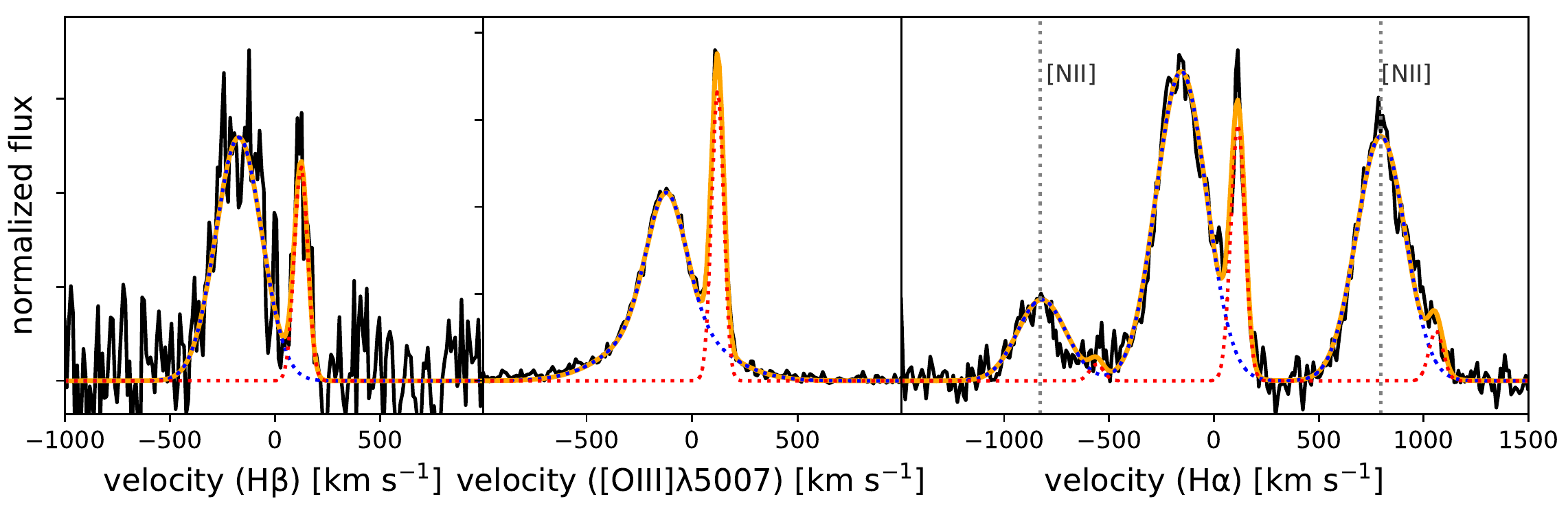}

\caption{Examples of \hb, \oiii\ and \ha\ velocity profiles at junction of ``inclined" structure and ``tail" structure. The red and blue dashed lines are the best-fitted spectra of red and blue peaks, respectively, with the sum of them in orange solid lines. The red and blue peaks reveal the ``tail" and ``inclined" structure, respectively.	
\label{fig:habo}}
\end{figure*}

To investigate the gas kinematics of the host galaxy through the 2-D profiles of narrow emission lines, we remove the emission from the quasar nucleus and the stars, each of which is represented by double Gaussians in the spatial regions where the NELs are negligible. The two top panels of Figure ~\ref{fig:oisii} show the restframe 2-D spectra in the neighborhood of \oi, \sii\ $\lambda\lambda$6717,6731 doublet, and \ha, \nii\ doublet, \hb, \oiii\ $\lambda\lambda$5007,4969 doublet (note that we employ the stellar redshift, $z = 0.22911$). The NEL structure extends to a size of $\sim$2.2$\arcsec$, greater than the FWHM of the point spread function (PSF) 0.71$\arcsec$ (see Appendix A), and thus is spatially resolved.

These 2-D NEL spectra reveal three components.
In particular, the 2-D spectra of \oi- and \sii-emitting NELs show a velocity gradient across the spatial extent, indicating a rotation-dominated disk. The velocity gradient can be seen in the 2-D NEL spectra produced using both the two independent exposures (see Appendix B), and thus is reliable. 
The 2-D spectra of \ha\ and \hb\ NELs (Figure ~\ref{fig:oisii}) imply the existence of an extended region located $\sim$ 4--20 kpc southwest to the quasar nucleus, whose position and velocity is reminiscent of extended emission line regions (EELRs) around quasars (e.g., \citealt{Stockton1987,Fu2009}).
The 2-D spectra of \oiii\ NEL is more complicated: besides the two components mentioned above, an additional blue-shifted component centered on the quasar nucleus is evident.
Such a profile is frequently seen in 2-D spectra of quasars' \oiii\ emission and is conventionally considered to be suggestive of the existence of outflowing gas. 
Therefore, the three components of the 2-D NEL spectra include a rotating disk, an EELR, and an outflow. In addition, we see double-peak profiles in  \oiii\, \ha\ and \hb\ lines in the Southwest of the nuclear, indicating multiple components.

\begin{figure*}[htb]
\centering
\includegraphics[width=0.48\textwidth]{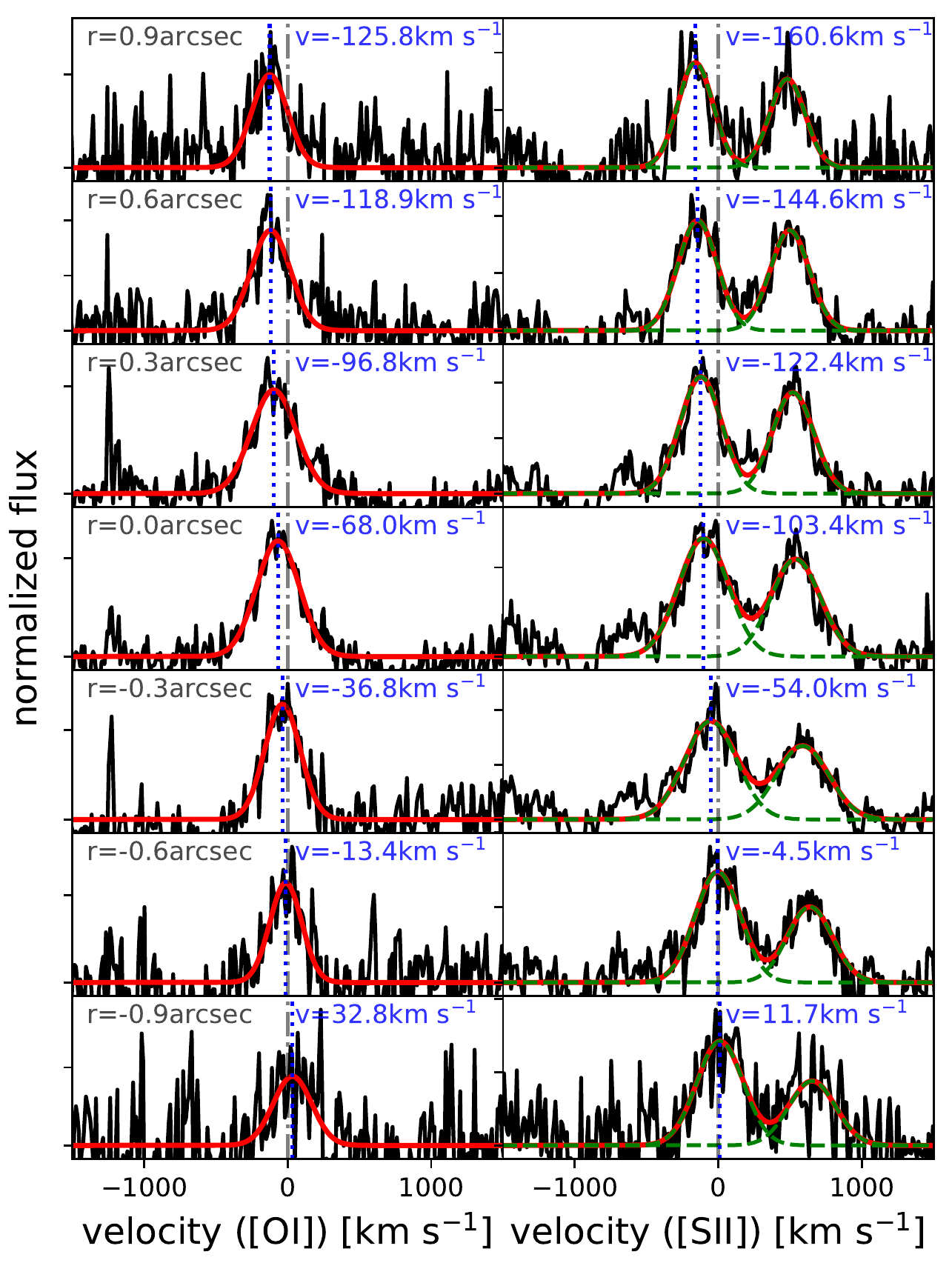}
\includegraphics[width=0.48\textwidth]{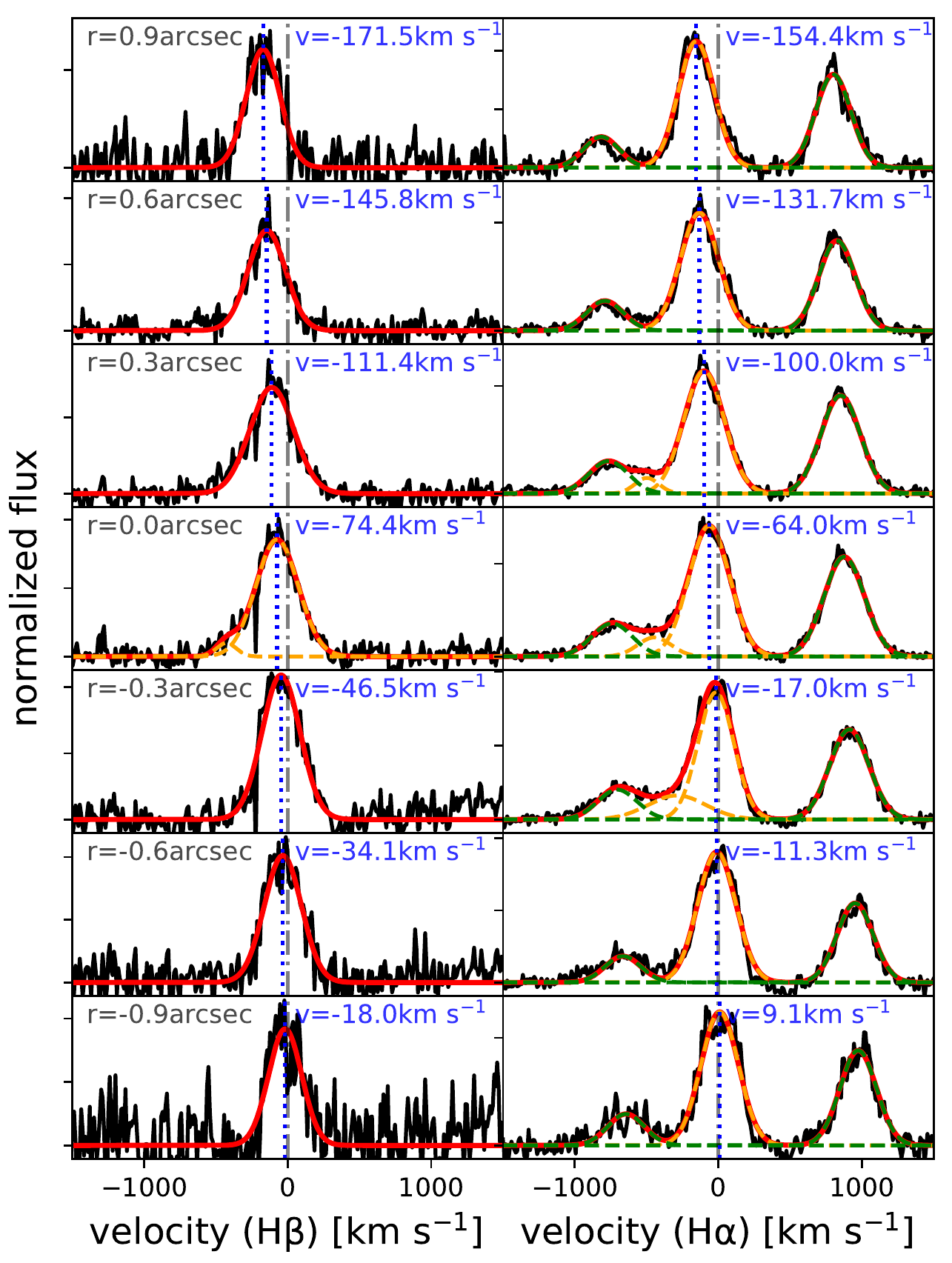}

\caption{The observed and best-fitted spectra of the \oi, \sii, \hb\ and \ha\ at different spaxels, from southwest to northeast from top to bottom. The project distance from the nuclear is shown on the upper right corner in each panel. The velocity of the rotating component is shown at the upper right corner in blue. For those spectra best fitted by a single gaussian, the median velocity is marked by a blue dotted line, and the zero point of the velocity which is measured by stellar component is in grey dashed dotted line. For those spectra best-fitted by two gaussians, the median velocity of the main gaussian (i.e., larger peak value) is marked by a blue dotted line. We note the \hb\ spectrum at the nuclear (r=0$\arcsec$) and \ha\ spectra within 0.3$\arcsec$ of the nuclear (r= 0.3, 0 and -0.3) are best fitted by two gaussians, due to the outflow component. The green dashed lines, in the second column, are the two gaussians from the double \sii\ lines. The orange dashed lines, in the third and fourth columns, are the two gaussians of \hb\ and \ha, respectively. The green dashed dotted lines, in the fourth column, are the best-fitted double \nii\ lines. 
\label{fig:rospec}}
\end{figure*}

To delineate the gas kinematics, we perform a two-step spectral fit to decomposition these three components using the Python package MPFIT. In the first step, for each spatial element (spaxel), we fit a double Gaussian to the profile of  \hb, \oiii, \ha, and \nii. The resultant best-fit Gaussian models for a single spaxel are demonstrated in Figure ~\ref{fig:habo}. Our reconstructed 2-D spectra of the red Gaussian show the EELR only, while the blue Gaussian is a superposition of a rotating disk and an outflow (Figure ~\ref{fig:oisii}, the third and fourth rows). In the second step, for each individual spaxel, we perform a single-Gaussian fit to the \oi\ emission line and the \sii\ $\lambda\lambda$6717,6731 doublet. The best-fit spectra of \oi\ and the \sii\ doublet in seven spaxels are shown in Figure ~\ref{fig:rospec} (left panels), from northeast to southwest. Due to the higher complication of their profiles, we fit the \ha\ and \hb\ profiles to a single or double Gaussian, for which the decision is made in a way similar to that in \citet{Liu2014}. It turns out that the 5 spectra closest to the nucleus demand for a double Gaussian. These fits allow us to successfully decompose the contribution from the rotating disk and a nearly spherical outflow structure, and the best-fit \ha\ and \hb\ spectra of seven spaxels are shown in Figure ~\ref{fig:rospec} (right panels), from northeast to southwest.


\subsection{The Rotating Disk}  \label{sec:disk}

\begin{figure}[htb]
\centering
\includegraphics[width=0.5\textwidth]{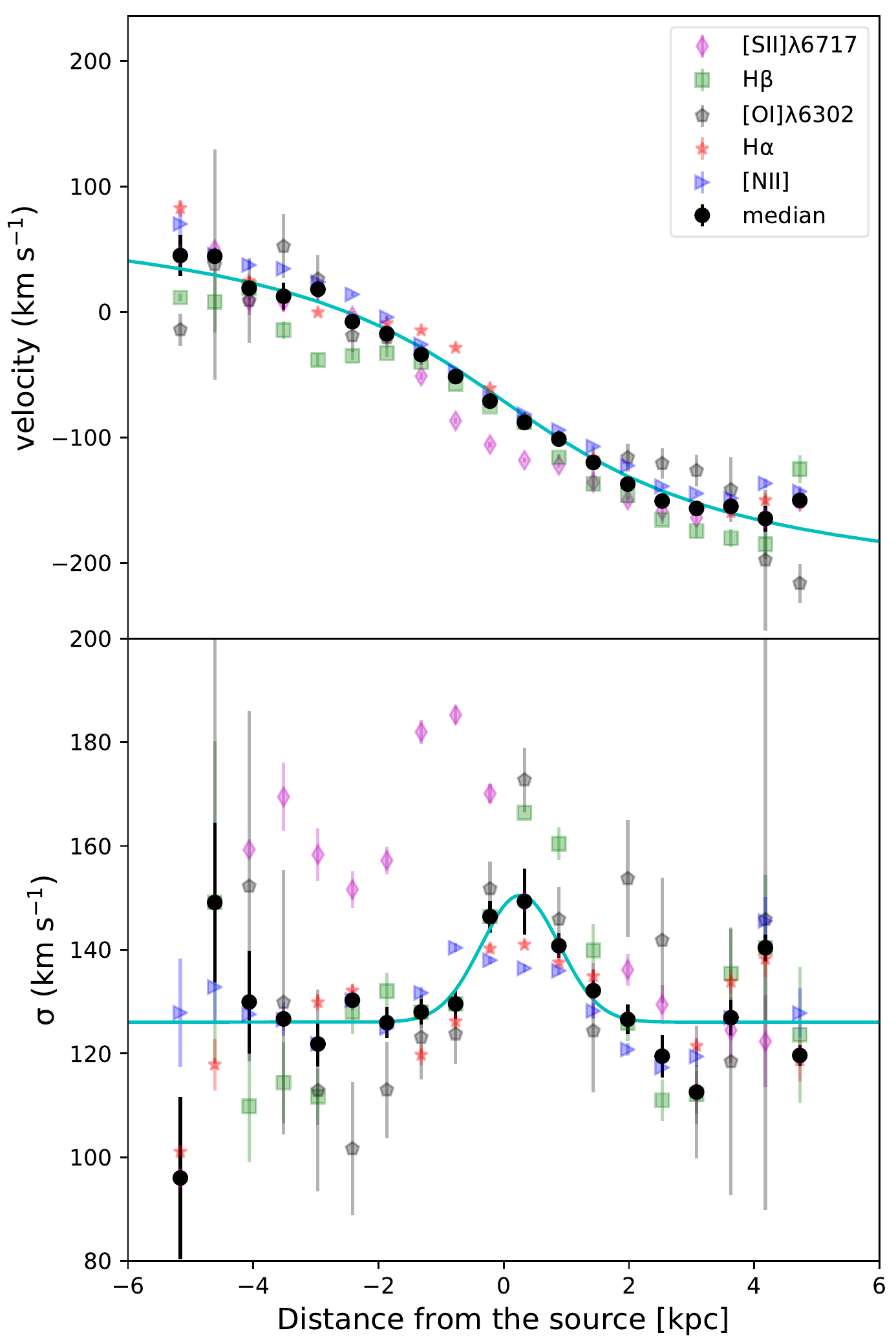}

\caption{Long-slit line-of-sight velocity and velocity dispersion profiles obtained with ESI/Keck.  The velocity (top) and velocity dispersion (bottom) of J0005+3820 (PA = -139$^{\circ}$) measured by \oi, \sii, \hb, \ha\ and \nii\ respectively are marked using grey pentagons, purple rhombus, green squares, red stars and blue triangles. And the black dots mark the median of the velocity and velocity dispersion measured by these emission lines. And the errors on median are calculated as $\sigma=\sigma_{\rm NMAD}/\sqrt{n - 1}$, where $\sigma_{NMAD}$ is the normalized median of the absolute deviations and n is to be 5 on behalf of the number of the sample.The cyan solid lines show the best fit of the data. The abscissa zero corresponds to the brightest-continuum bin along the slit (to the galactic nucleus).
\label{fig:rovw}}
\end{figure}

We scrutinize the rotating disk by measuring the velocity relative to the redshift of host galaxy and the velocity dispersions of best-fit spectra of the rotation components in the NELs of \oi, \sii, \hb, \ha\ and \nii. In Figure ~\ref{fig:rovw}, we plot them as a function of the distance from the quasar nucleus (positive is to the southwest and negative to the northeast). In addition, we only show those spaxels with high signal to noise ratio (S/N of \oi\ $\gtrsim$ 3), corresponding to $\sim$ 5 kpc from the nucleus.
We see that the overall velocity distribution and dispersion of these NELs are in line with typical galactic disks \citep{Rix1992,Proshina2020}. We find that the velocity profiles measured from different NELs are consistent, and thus use the median of the measurement results from these different NELs, following the method employed by \citet{Courteau1997}, \citet{Weiner2006} and \citet{Drew2018}. In detail, assuming an axis ratio of $b/a = 1$, the distribution of velocity is formulated as:
\begin{equation}
 V_{\rm obs}(r) = (\frac{2}{\pi}V_{\rm a}\arctan(\frac{r_{\rm obs}}{r_{\rm t}\cos(i)}) + c\frac{r_{\rm obs}}{\cos(i)})  \times \sin(i) + V_{\rm rel},
\end{equation}
where $V_{\rm a}$ is the asymptotic velocity, $r_{t}$ is the knee radius, $r_{\rm obs}$ is the projection distance from the center of the galaxy, $c$ is the outer-galaxy slop, and $V_{\rm rel}$ is the velocity of the ionized gas relative to the stars along the line of sight, and $i$ is the inclination angle of the galaxy. Here $V_{\rm a}$ and $r_{\rm t}$ are dimensional scaling parameters, whereas $c$ characterizes the shape of the rotation curve.
Meanwhile, if a Gaussian fit is used, the profile of the velocity dispersion is given by:
\begin{equation}
\Sigma(r) = A {\rm exp}(-\frac{(r-m)^{2}}{2\omega^{2}} + \sigma_{0}),
\end{equation}
where $\Sigma(r)$ is the value of the fitted Gaussian at each radius, $m$ is the central position (if the ionized gas is isotropically distributed around the center, then $m$ is 0), $\omega$ is the width of the Gaussian, and $\sigma_{0}$ is the isotroptic component of the velocity dispersion. Here we correct the observed velocity dispersion for the intrinsic instrmental dispersion using $\sigma = \sqrt{\sigma_{\rm obs}^{2} - \sigma_{\rm inst}^{2}}$, where $\sigma$ reported in Figure ~\ref{fig:rovw}, and $\sigma_{\rm inst}$ is the combined instrumental and spectral seeing dispersion that we measure to be $\sim$ 23 km s$^{-1}$. The resultant best-fit velocity dispersion $\sigma$ is 126 $\pm$ 10 km s$^{-1}$.

We fit the velocity and velocity dispersion profiles to the median values of the measurements results from different NELs (\oi, \sii, \hb, \ha\ and \nii), utilizing a Monte-Carlo simulation that we run 50 times to calculate the errors of parameters. Errors on the median are calculated as  $\sigma=\sigma_{\rm NMAD}/\sqrt{n - 1}$, where $\sigma_{\rm NMAD}$ is the normalized median of the absolute deviations \citep{Hoaglin1983} and $n$ is the number of the sample, here $n = 5$. The best-fit profiles are shown in Figure ~\ref{fig:rovw}, along with the corresponding parameters listed in Table ~\ref{tab:tab3}. The velocity profile reveals a roughly symmetric gas motion pattern. Gas in the quasar's nuclear region has a velocity of $\sim$ -75 km s$^{-1}$, and is blue-/red-shifted on the SW/NE side, respectively. The velocity profile flattens beyond a radius of 4 kpc in both of these two directions, with a velocity of -150 to -180 km s$^{-1}$ (SW) and $\sim$ 40 km s$^{-1}$ (NE) within the distance range of 4 to 6 kpc. The rotating component is blueshifted relative to the systemic (stellar) velocity with $V_{\rm rel} =$ -71 km s$^{-1}$. Since the narrow absorption line system is blue-shifted with a similar relative velocity (-85 km s$^{-1}$; see Sections \ref{sec:nabsya}), this kinetic consistency is suggestive of the same origin of the two.

In the central region, we note that a velocity difference of $\sim$ 50 km s$^{-1}$ exists at a $> 3\sigma$ significant level between \sii\ and other NELs. This is likely due to the outflow, though the possibility of influence from the sky emission line in the blue outskirt of \sii\ cannot be fully ruled out; but if \sii\ is excluded from these analyses, our results change minimally.

\begin{table}[htb]\footnotesize
\caption{Parameters for modelling the velocity shift and velocity dispersion.}

\begin{tabular}{ccc}
\hline
\hline
Parameter   & Value & Unit\\
\hline
\multicolumn{3}{c}{velocity shift}\\
\hline
$V_{\rm a}$                & $-196\pm48$ & km s$^{-1}$ \\
$r_{\rm t}$          & $4.1\pm1.1$ & kpc \\
$c $     & $-0.8\pm4.8$  \\
$V_{\rm rel} $     & $-71\pm4$ & km s$^{-1}$ \\
$i$     & $0.84\pm0.07$ & rad \\
\hline
\multicolumn{3}{c}{velocity dispersion} \\
\hline
$A$  & $24\pm2$  \\
$m$           & $0.26\pm0.01$ & kpc \\
$\omega$           & $0.63\pm0.06$ & kpc \\
$\sigma_{0}$           & $126\pm3$ & km s$^{-1}$ \\
\hline
\end{tabular}

\label{tab:tab3}
\end{table}


\subsection{The Extended Emission Line Region}  \label{sec:eelr}

\begin{figure*}[htb]
\centering

\includegraphics[width=0.9\textwidth]{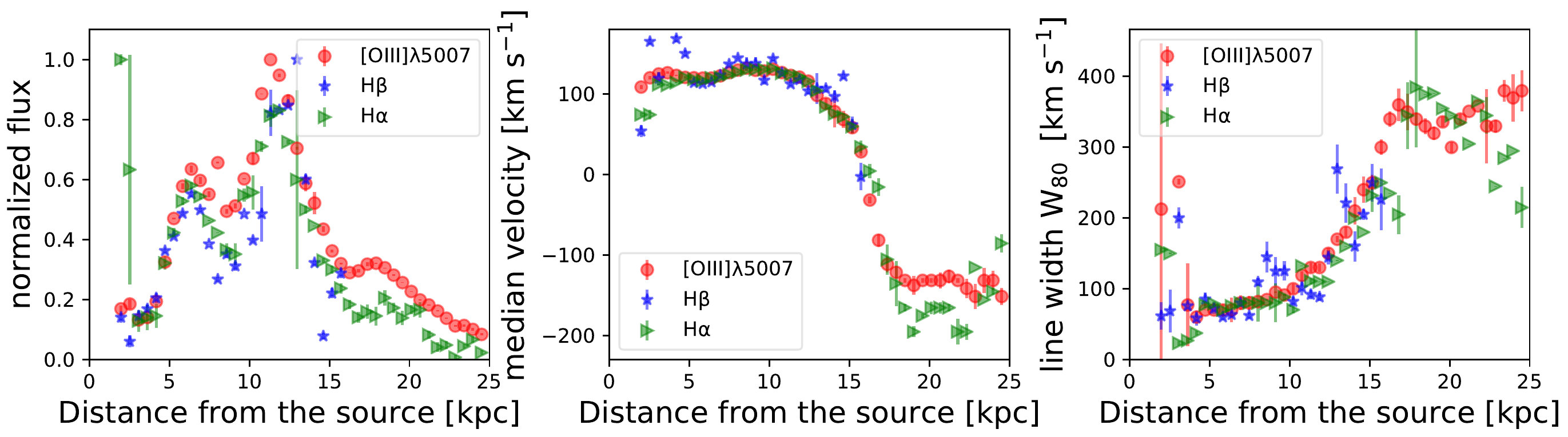}
\caption{Long-slit line-of-sight velocity, velocity dispersion and normalized flux density profiles obtained with ESI/Keck. The velocity, velocity dispersion and normalized flux measured by \oiii, \hb\ and \ha\  respectively are marked using red, yellow and blue. 
\label{fig:ep}}
\end{figure*}

The EELR extends to the southwest of the nucleus (see Figure~\ref{fig:oisii}), which is most evident in the \oiii\ emission, where two compact knots and a diffuse region are seen. For the line spectra at each position, we measure the integrated flux, median velocity, and line width ($W_{80}$, as detailed in \citealt{Liu2013b}) of \ha, \hb\ and \oiii\ emission lines by fitting their profiles to Gaussian models, and conclude that the three NELs depict consistent velocity profiles (Figure ~\ref{fig:ep}).

\begin{figure}[htb]
\centering
\includegraphics[width=0.45\textwidth]{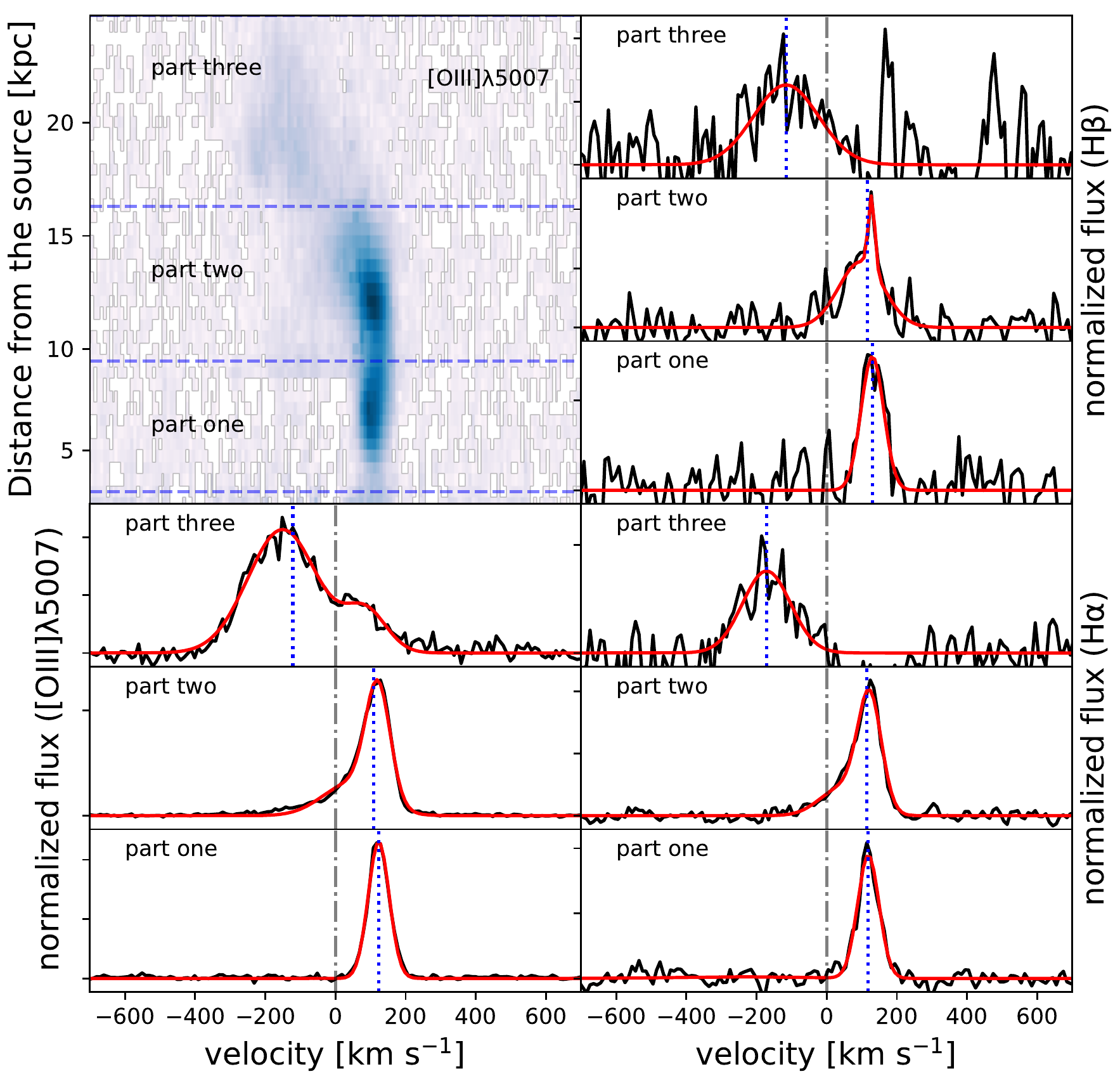}

\caption{Spectra of the \ha, \hb\ and \oiii at three different regions. Red line shows the best fitting, and the blue dotted line marks the median velocity, the velocity zero is in grey dashed dotted line.\label{fig:esp}}
\end{figure}

The integrated flux profile of all three NELs unambiguously show a series of three peaks at galactocentric distances of about 7, 13 and 18 kpc. Hence, we divide the EELR into three annuli, with a radius range of 2 - 8 kpc, 8 - 16 kpc and 16 - 25 kpc, respectively, and for each of which, we extract a 1-D emission line spectra (Figure ~\ref{fig:esp}). The best-fit integrated fluxes, median velocities and line widths are listed in Table~\ref{tab:tab4}, where it can be seen that the emission lines from the 2 - 16 kpc are relatively narrow and are redshifted relative to the stellar redshift, while at 16 - 25 kpc they are broader and blue-shifted.

\begin{figure}[htb]
\centering
\includegraphics[width=0.45\textwidth]{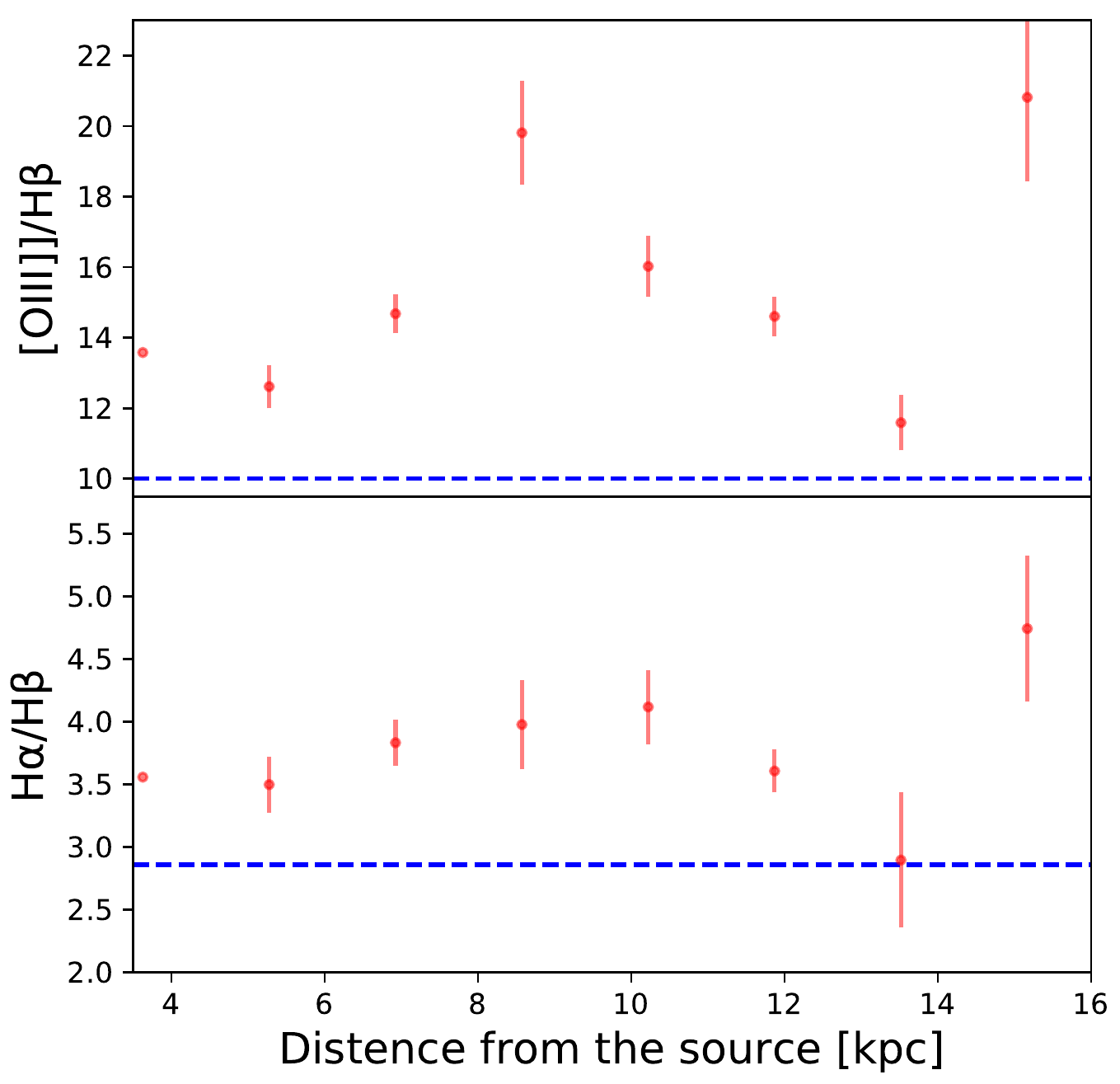}

\caption{Line ratio of the \oiii/\hb\ and \ha/\hb\ of the extended ionized gas profiles along the long-slit line-of-sight. Blue dash lines mark the typical \oiii/\hb\ ratio and ``Case B" value, respectively. 
\label{fig:er}}
\end{figure}

The intensity ratios of the emission lines facilitate our analysis on the physical conditions of the EELR. In specific, we use the \oiii/\hb\ and \ha/\hb\ ratios to quantify the degree of ionization and the dust attenuation, respectively. To obtain higher S/N ratios, we bin the spatial pixels within 0.45 arcsec before measuring the median values of these line ratios, for which the uncertainties are, again, calculated using the MC method (Figure ~\ref{fig:er}). We find that  \oiii/\hb\ is larger than 10 at all galactocentric radii under our consideration (top panel therein), implying for a high-ionization state in general. Meanwhile, considering the theoretical ratio \ha/\hb\ $\sim$ 2.86 for ``case B" ($n_{e}$ = 100 cm$^{-3}$, $T_{e}$ = 10000 K;\citealt{Storey1995}), we plot the result in Figure ~\ref{fig:er} (blue dash line, bottom panel). The \ha/\hb\ ratio at galactocentric radii of 2 - 16 kpc are roughly constant, corresponding to $A_V$ $\sim$1. However beyond 16 kpc, \hb\ is too weak for accurate measurement of dust attenuation.

\begin{figure*}[htb]
\centering
\includegraphics[width=0.9\textwidth]{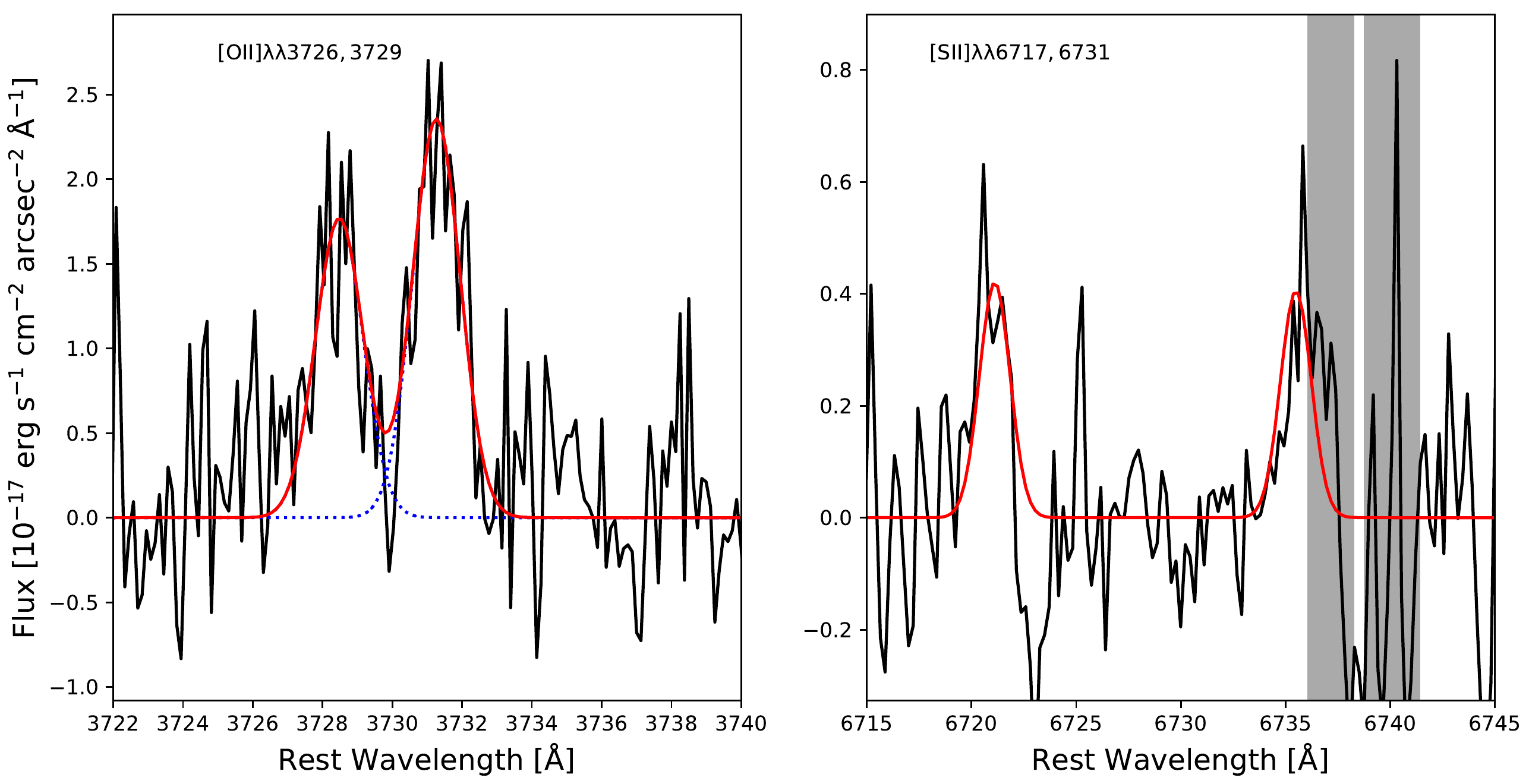}
\caption{Spectra of the \oii\ and \sii\ emission line in the rest frame. The fitted line is in red. The blue dashed line shows the two gaussians from the double \oii\ lines. The grey shaded region are influenced by sky line.
\label{fig:eoii}}
\end{figure*}

The electron density is achievable from \oii\ $I(3729)/I(3726)$ and \sii\ $I(6717)/I(6731)$ ratios. For this purpose, we stack the spectra taken from locations 2-16 kpc away from the center, and fit the [O ii] and [S ii] doublet emission lines by fixing the kinematics of the two line (Figure ~\ref{fig:eoii}). Assuming an electron temperature of 10000K and at a  68\% confidence level, we estimate the electron density  to be $n_{e}$ = 137 $^{+36}_{-30}$ cm$^{-3}$. This result is generally lower than that of typical narrow line regions ($n_{e}$ $\sim$ 1000 cm$^{-3}$;\citealt{Greene2011}), but close to that of EELR \citep{Fu2009}.

\begin{table}[htb]\footnotesize
\caption{Nonparametric measurements  of three parts.}

\begin{tabular}{ccc}
\hline
\hline
Parameter   & Value & Unit\\
\hline
\multicolumn{3}{c}{part one}\\
\hline
$f_{\oiii}$          & $28.8\pm0.10$ & 10$^{-17}$ erg s$^{-1}$ cm$^{-2}$ arcsec$^{-2}$ \\
$v_{\oiii}$                & $123\pm0.12$ & km s$^{-1}$ \\
$W_{\oiii}$          & $77\pm0.31$ & km s$^{-1}$ \\
$f_{\ha}$          & $8.6\pm0.32$ & 10$^{-17}$ erg s$^{-1}$ cm$^{-2}$ arcsec$^{-2}$ \\
$v_{\ha}$                & $119\pm0.61$ & km s$^{-1}$ \\
$W_{\ha}$          & $79\pm1.58$ & km s$^{-1}$ \\
$f_{\hb}$          & $2.1\pm0.12$ & 10$^{-17}$ erg s$^{-1}$ cm$^{-2}$ arcsec$^{-2}$ \\
$v_{\hb}$                & $130\pm2.11$ & km s$^{-1}$ \\
$W_{\hb}$          & $82\pm5.82$ & km s$^{-1}$ \\
$A_{V}$             & $1.186\pm0.233$ & mag \\
\hline
\multicolumn{3}{c}{part two} \\
\hline
$f_{\oiii}$          & $52.6\pm0.72$ & 10$^{-17}$ erg s$^{-1}$ cm$^{-2}$ arcsec$^{-2}$ \\
$v_{\oiii}$                & $108\pm2.33$ & km s$^{-1}$ \\
$W_{\oiii}$          & $170\pm3.03$ & km s$^{-1}$ \\
$f_{\ha}$          & $12.3\pm1.60$ & 10$^{-17}$ erg s$^{-1}$ cm$^{-2}$ arcsec$^{-2}$ \\
$v_{\ha}$                & $114\pm6.75$ & km s$^{-1}$ \\
$W_{\ha}$          & $149\pm8.44$ & km s$^{-1}$ \\
$f_{\hb}$          & $3.4\pm0.17$ & 10$^{-17}$ erg s$^{-1}$ cm$^{-2}$ arcsec$^{-2}$ \\
$v_{\hb}$                & $117\pm10.08$ & km s$^{-1}$ \\
$W_{\hb}$          & $150\pm25.15$ & km s$^{-1}$ \\
$A_{V}$            & $0.860\pm0.475$ & mag \\
\hline
\multicolumn{3}{c}{part three} \\
\hline
$f_{\oiii}$          & $21.3\pm0.23$ & 10$^{-17}$ erg s$^{-1}$ cm$^{-2}$ arcsec$^{-2}$ \\
$v_{\oiii}$                & $-122\pm7.75$ & km s$^{-1}$ \\
$W_{\oiii}$          & $348\pm8.38$ & km s$^{-1}$ \\
$f_{\ha}$          & $3.3\pm1.70$ & 10$^{-17}$ erg s$^{-1}$ cm$^{-2}$ arcsec$^{-2}$ \\
$v_{\ha}$                & $-171\pm4.59$ & km s$^{-1}$ \\
$W_{\ha}$          & $175\pm11.87$ & km s$^{-1}$ \\
$f_{\hb}$          & $2.4\pm0.16$ & 10$^{-17}$ erg s$^{-1}$ cm$^{-2}$ arcsec$^{-2}$ \\
$v_{\hb}$                & $-116\pm6.56$ & km s$^{-1}$ \\
$W_{\hb}$          & $238\pm17.35$ & km s$^{-1}$ \\

\hline
\end{tabular}
\tablecomments{Flux of the \ha, \hb\ and \oiii\ are directly measured, not do dust attenuation correction. The velocity is the median velocity measured by \ha\, \hb\ and \oiii, and $W$ is the $W_{80}$\citep{Liu2013b}. We use the Milky Way attenuation curve starbust galaxies and the average extinction \citep{Calzetti2000} to reddening relation at the $V$ band of $A_{V} = 4.05E(B-V)$}
\label{tab:tab4}
\end{table}

\section{Discussion} 
\label{sec:discussion}

\subsection{host galaxy}  \label{sec:dr}

As we have introduced in the introduction section, there were abundant works studying the host galaxies of blazars, particularly to distinguish whether the host galaxies are disk or elliptical galaxies. Most of them focused on imaging BL Lac, while, none of them successfully found a disk galaxy hosted FSRQ. Because blazars are rare and can only be found in the distant universe, even using HST with the highest spatial resolution in optical and NIR bands, cannot spatially decompose the nuclei and the host galaxies well. However, we could use spatially resolved spectrum to constrain the host galaxy properties. In disk galaxies, young stars and interstellar gas and dust rotate in disks around bulging nuclei, while in elliptical galaxies, old stars randomly swarm and gas and dust are lack. Thus, the properties and kinematics of stars and gases can be used to distinguish between disk and elliptical galaxies.

In Section \ref{sec:disk}, we find that the kinematics of gas in the host galaxy of B2 0003+38A is dominated by rotating, and the curvatures of velocity and velocity dispersion are similar with those of disk galaxies \citep{Ho2020}.
In addition, there are other hints that supports B2 0003+38A being hosted by a disk galaxy. The velocity of the ionized gas (rotating component) relative to the systemic (stellar) velocity with $V_{\rm rel} =$ -71 km s$^{-1}$ is similar with the velocity of the narrow absorption line system (-85 km s$^{-1}$), suggesting the same origin of these two. Furthermore, emission lines from host galaxy can be detected significantly, indicating a gas-rich host galaxy. 
The spectrum lacks the features of old stars, such as TiO molecular bands, indicating that old stars are not the dominant. Meanwhile, the spectrum is red, suggesting that the starlight is dust reddened. Young stars and rich interstellar dust are also characteristic of disk galaxies.
This is consistent with the analysis of nuclear spectrum prefers young stellar population ($\sim$450 Myr) with large dust reddening ($E(B-V)_\star=0.62$), under the assumption of a single SSP and quasar spectrum. Therefore, we conclude that the B2 0003+38A is mostly likely hosted by a disk galaxy.

\subsection{The Extended Emission Line Region}  \label{sec:dr}

We find the EELR in southwest of the nucleus that extends to a projected distance up to 25 kpc.
EELRs with such sizes were found around a substantial fraction of radio-loud quasars.

Before further discussing the origin of the EELR, we estimate some parameters of the EELR.
The gases in part one and part two have redshifted velocities of $v_{0} \sim 120$ km s$^{-1}$, and their distance to the galaxy nucleus is $R_{0} \sim 15$ kpc (see  Figure ~\ref{fig:esp}).
We estimate a dynamical time scale to be $t \sim 1.2 \times 10^{8}$ yr, i.e. the time taken by the gas from the nucleus to reach such a distance with an average velocity of $v_{0}$.
The mass of gas in the two parts can be estimated using \hb\ luminosity $L_{\hb}$ and electron density $n_{e}$ (e.g \citealt{Liu2013,Harrison2014}).
After corrected for dust attenuation, the summed luminosity in the two parts is $\sim$ 2.4$\times10^{40}$ erg s$^{-1}$.
The total mass of these ionized extended gas can be estimated as:
\begin{equation}
 \frac{M_{\rm gas}}{2.82 \times 10^{9} \rm ~ M_\odot} = (\frac{L_{\hb}} {10^{43} \rm ~ erg ~ s^{-1}}) (\frac{n_e} {100 \rm ~ cm^{-3}}).
\end{equation}

We measure an electron density of $n_{e}$ $\sim$ 137 cm$^{-3}$ using \oii\ and \sii\ doublet, if the electron temperature is 10000K.
We find $M_{\rm gas} \sim 9.3 \times 10^{6}$ $M_\odot$.
Combining with the average velocity of 120 km s$^{-1}$, the total kinetic energy of the gas can be estimated as:
\begin{equation}
E_{\rm kin} = \frac{1}{2}M_{\rm gas} v_{\rm gas}^{2} = 1.33 \times 10^{54} \rm ~ erg.
\end{equation}
If assuming the lifetime of the structure is the dynamical time scale, we can also estimate a mass rate $\dot{M}$ is  7.8 $\times 10^{-2}$ M$_\odot$ yr$^{-1}$, and a kinetic energy rate $\dot{E}_{\rm ink} \sim$ 3.5 $\times 10^{38}$ erg s$^{-1}$.

However, this electron density value may not be the bulk density of the EELR gas.
The EELR might be illuminated by the quasar.
If so, the ionization parameter $U$ of the EELR can be estimated as:
\begin{equation}
U = \frac{ Q_H }{ 4\pi r^2 c n_H },
\end{equation}
where $Q(H)$ is the rate of the hydrogen ionization photon from the quasar, $r$ is the distance from the quasar nucleus to the EELR, $n_H$ is the hydrogen number density.
Assuming that the intrinsic SED of the quasar is that given in \citet{Mathews1987}, we estimate a $Q_H$ of $1.2\times10^{56}$ s$^{-1}$ using the bolometric luminosity estimated in Section \ref{sec:2ds}.
Assuming a distance of 10 kpc, and assuming $n_e=1.2 n_H$ for highly ionized plasma, the inferred $U$ is about $10^{-2.5}$.
However, the observed \oiii/\hb\ values of 12--16 indicates a higher $U$ value of $\sim10^{-1}$.
This may be because the \oiii\ emitting gas has a lower density than previously estimated value. 
Using the mixed-medium model \citep{Robinson2000}, \cite{Stockton2002} divided the EELR of 4C 37.43 into two components, and found that one component with a density of several hundred cm$^{-3}$, another with a density near 1 cm$^{-3}$.
Most of the \oiii, \hb\ and \ha\ come from regions with low density, \oii\ and \sii\ come from the regions with density several hundred times higher. If this is the case for the EELR in B2 0003+38A, actual value of $M_{\rm gas}$ is two orders of magnitude higher.

There are various possibilities for the origin of an EELR.
In the case of an isolated galaxy, the gas can be inflow, i.e. cold accretion flow from the intergalactic medium, outflow triggered by AGN or starburst, or recycling gas (of the outflow).
In the case of galaxy with interactions, the gas can also belong to tidal features.
Firstly, inflows are generally isotropic, which excludes the possible of EELR being inflow driven.
Secondly, EELRs are common in radio quasars and these EELRs are generally related with outflows driven by radio jets (e.g., \citealt{Fu2009}), so the EELR in B2 0003+38A may also be the same. Thirdly, verifying the possibility of recycling gas requires a panoramic image of the circum-galactic gas.
The long-slit spectrum is not sufficient, and future Integral Field Unit (IFU) observation data is needed.
We did not see any companion galaxy around B2 0003+38A, and did neither see any asymmetry from the brightness profile of the starlight. As for the possible of galaxy interactions, we do not find any direct evidence including any signature of asymmetry from the brightness profile of the starlight and any companion galaxy around it.
Though, we note that the current data quality is not enough to rule out the possibility.
The EELR may also originate in tidal features.
If so, starlight accompanied with EELR should be seen.
This cannot be tested using the existing data. Future observations, such as IFU, with higher quality are required to fully investigate the origin of EELR.

\section{Conclusions} 
\label{sec:conclusions}

In this paper, we present a long-slit observations taken from the ESI/Keck to study the gas in the host galaxy of FSRQ B2 0003+38A at redshift $z =$ 0.22911. Based on multiple Gaussian fitting processes, we separate the 2-D NEL spectra into three components, indicating a rotation disk, an EELR and an outflow, respectively. To model the rotating disk, we measure and analyse the curves of velocity and velocity dispersion. We also analyse the EELR, which extends to a projected distance up to 25 kpc from the nuclear. We summarize our results below.

(i) For the first time, we discover a rotating gasous disk from optical spectroscopy in a FSRQ host galaxy. The curvatures of velocities and the velocity dispersions derived from different emission lines agree with an identical rotating disk model. The rotating gas disk has a mean velocity of $v$ = $-$75 km s$^{-1}$ relative to the stellar redshift.  The velocity has little difference with that of the absorber seen in Na I, Ca II, and H I lines ($v$ = $-$85 km s$^{-1}$), suggesting that the rotating gas disk and the absorber are related.

(ii) According to the kinematics and morphology of the EELR, we divide them into three parts, including two knots and a diffused region. We calculate that the two knots have an averaged electron density $n_{e}$ of 137 $^{+36}_{-30}$ cm$^{-3}$ and dust attenuation $A_{V}$ of 1.19 $\pm$ 0.23/0.86 $\pm$ 0.48, respectively. After correcting for the dust attenuation, we estimate a corresponding mass to be 9.3 $\times 10^{6}$ M$_\odot$. The velocity of ionized gas in these two knots is redshifted 120 km s$^{-1}$. The dynamical timescale of the knots can be estimated at $\sim 1.2 \times 10^{8}$ yr as the travel time of clouds to reach the observed distances from the centre. There are various possibilities for the origin of an EELR, including inflows, outflows, recycling gas and galaxy interactions. If the EELR is from an outflow/inflow, the mass rate is 7.8 $\times 10^{-2}$ M$_\odot$ yr$^{-1}$, and the kinetic energy carried by ionized gas is estimated as 1.46 $\times 10^{54}$ erg.

\acknowledgements

QZ, LS and GL acknowledge the grant from the National Key R\&D Program of China (2016YFA0400702), the National Natural Science Foundation of China (No. 11673020 and No. 11421303), and the Fundamental Research Funds for the Central Universities. We acknowledge the support from Chinese Space Station Telescope (CSST) Project. 

\begin{appendices}
\section{Measuring the Point Spread Function}
\begin{figure}[htb]
\centering
\includegraphics[width=0.45\textwidth]{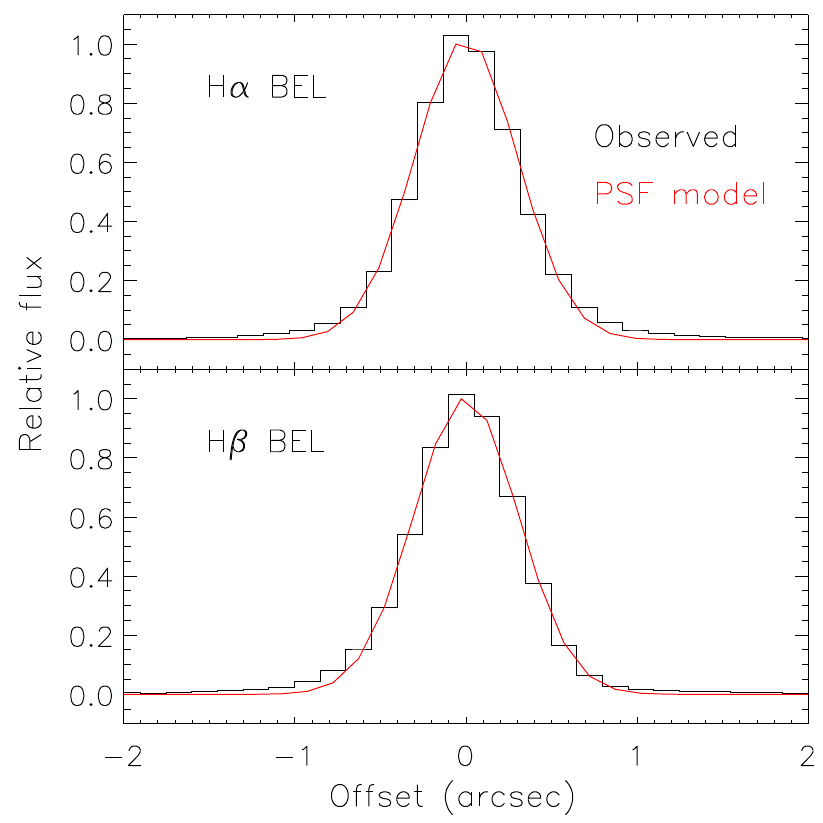}

\caption{The spatial brightness profiles of H$\alpha$ and H$\beta$ BELs, and the Gaussian functions that fit them.
\label{fig:psf}}
\end{figure}

We measure the PSF due to seeing effect and instrumental effect using the spatial brightness profiles of BELs.
We can do in this way because the BELR, with a typical size less than 1 pc, is a point source at $z\sim0.2$.
We extract the spatial brightness profiles of the H$\alpha$ and H$\beta$ BELs (Figure ~\ref{fig:psf}.
Both the two profiles can be well fit using a Gaussian function with a FWHM of 0.71$\arcsec$.

\section{Verifying the velocity gradient of NELs across the spatial extent}

\begin{figure*}[htb]
\centering
\includegraphics[width=0.9\textwidth]{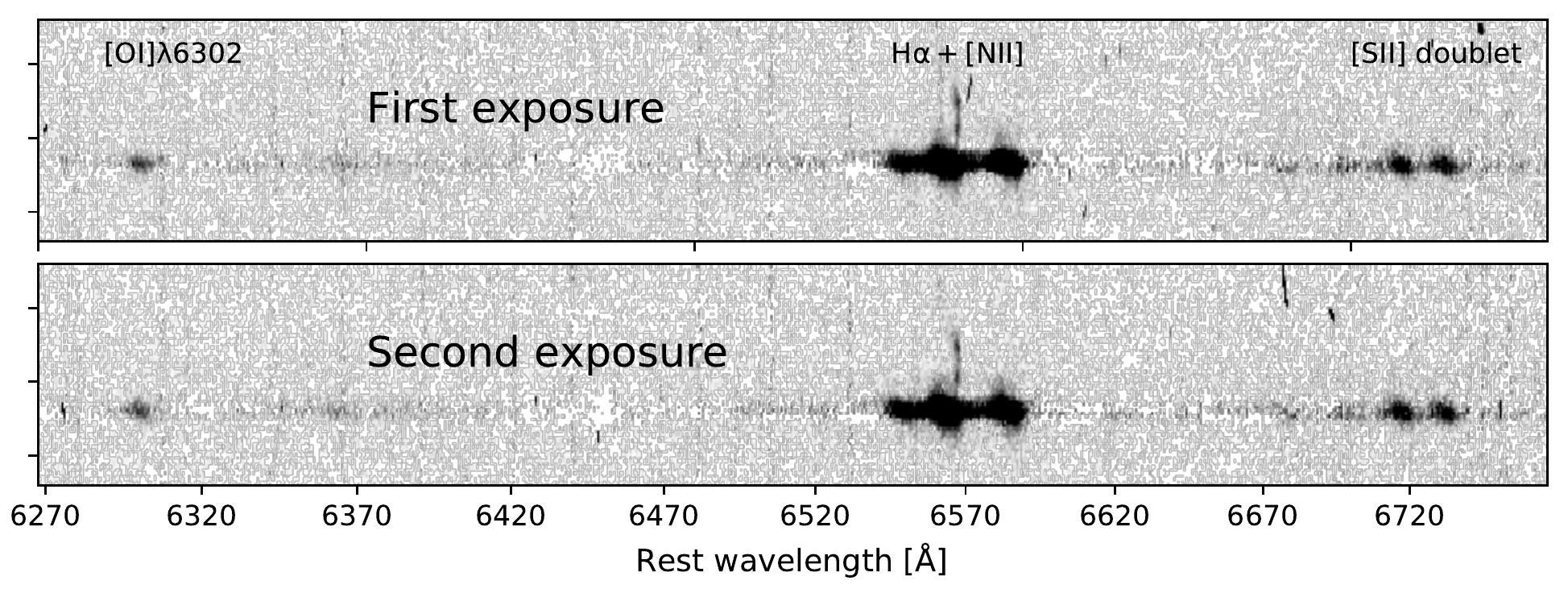}

\caption{Same as Figure ~\ref{fig:oisii}. First row: the observed 2-D spectra of the \oi, \ha, \nii\ and \sii\ from the data of the first exposure, second row: those from the second exposure. Note that the cosmic rays are not removed. 
\label{fig:ctwo}}
\end{figure*}

One may doubt that velocity gradient of NELs across the spatial extent seen from the 2-D spectra might be artificial: it might be caused by a tiny inclined movement of the target along the slit during the observation.
As there were two exposures of this target, we test this presumption by independently analyzing the data from the two exposures.
We reduce the data and extract the 2-D NEL spectra again following the methods described in Sections \ref{sec:dar} and \ref{sec:2dsnl}, while this time we do not stack the two exposures.
The results are shown in Figure ~\ref{fig:ctwo}.
The velocity gradient of NELs are seen in both the two 2-D NEL spectra, indicating that the rotation-dominated disk structure in B2 0003+38A is reliable.

\end{appendices}




\bibliography{./wenxian.bib}


\end{document}